\documentstyle[preprint,aps]{revtex}
\begin{document}
\title{Self-induced and induced transparencies of two-dimensional and three-
dimensional superlattices}
\author{Yuriy A. Romanov and Julia Yu. Romanova}
\address{Institute for Physics of Microstructures RAS, 603600 Nizhny Novgorod, \\
Russia}
\author{Lev G. Mourokh and Norman J.M. Horing}
\address{Department of Physics and Engineering Physics, \\
Stevens Institute of Technology, Hoboken, NJ 07030 }
\date{\today }
\maketitle

\begin{abstract}
{\ The phenomenon of transparency in two-dimensional and three-dimensional
superlattices is analyzed on the basis of the Boltzmann equation with a
collision term encompassing three distinct scattering mechanisms (elastic,
inelastic and electron-electron) in terms of three corresponding distinct
relaxation times. On this basis, we show that electron heating in the plane
perpendicular to the current direction drastically changes the conditions
for the occurrence of self-induced transparency in the superlattice. In
particular, it leads to an additional modulation of the current amplitudes
excited by an applied biharmonic electric field with harmonic components
polarized in orthogonal directions. Furthermore, we show that self-induced
transparency and dynamic localization are different phenomena with different
physical origins, displaced in time from each other, and, in general, they
arise at different electric fields.}
\end{abstract}


\narrowtext
\newpage

{\large {\bf I. Introduction} \vspace*{1cm} }

{\normalsize Semiconductor superlattices have been at the focus of attention
for several decades, due to their unique electronic properties. The
additional spatial periodicity of the superlattice leads to the formation of
narrow Brillouin minizones and energy minibands \cite{1,2,3}. Bloch
oscillations \cite{4} and Wannier-Stark levels \cite{5} can be observed in
superlattices due to the narrowness of these minibands even in relatively
weak static electric fields ($10^{2}-10^{4}V/cm$). The Bloch oscillations
are due to Bragg reflections by the periodic superlattice potential and are
characterized by the frequency $\Omega _{c}=eE_{c}d/\hbar $ and amplitude $%
Z_{c}=\Delta /2eE_{c}$, where $E_{c}$ is the constant electric field applied
along the axis of the superlattice of period $d$ and miniband width $\Delta $%
. In the case of an applied harmonic {\it ac} field, the Bragg reflections
do not generate a new type of oscillation beyond that of the static field,
but they do modulate electron motion during the field period. This
modulation is described by oscillatory dependencies of the amplitudes of
electron velocity harmonics on the amplitude, $E_{1}$, and/or the frequency, 
$\omega _{1}$, of the applied harmonic field \cite{22,23}. Manifestations of
this modulation can be found in various nonlinear macroscopic effects and,
in particular, in {\it superlattice transparency} \cite{23,16,24,25,26}. The
zero$^{th}$ harmonic of electron nonlinear oscillations responsible for {\it %
dc} current is of special interest. Its vanishing corresponds to electron
spatial localization and is called {\it dynamic localization} \cite{22}.
This dynamic localization occurs only for electrons having a sine-like
dispersion law and for specific ratios of amplitude and frequency of the
applied field, such that $J_{0}(eE_{1}d/\hbar \omega _{1})=0$ (where $%
J_{0}(x)$ is the zero$^{th}$ order Bessel function). In the case of
deviation from a sinusoidal dispersion law, dynamic localization can arise
only at multifrequency fields \cite{23}. In the literature (see, for
example, Ref.\cite{???}), dynamic localization is very often erroneously
identified with self-induced superlattice transparency, predicted in Ref.
\cite{24} and verified experimentally in Ref.\cite{25}. It was shown there
that the macroscopic polarization of the electron gas can vanish and the
superlattice behaves almost like a linear dielectric having the permittivity
of the crystal lattice in the absence of mobile electrons, with small
nonlinear absorption due to electron oscillations. The conditions for this
effect within the single $\tau $-approximation for a one-dimensional
superlattice sample are the same as for dynamic localization, but they have
different physical origins. The error of their identification was shown in
Refs. \cite{23,27} for a three-dimensional sample with a one-dimensional
superlattice and it will be further confirmed below for two- and
three-dimensional superlattices. }

{\normalsize The single $\tau$-approximation has been reasonably successful
in describing the cases of a one-dimensional superlattice and also a
one-dimensional model of a three-dimensional superlattice {\it without}
redistribution of energy and momentum among the various degrees of freedom
due to electron scattering. However, as was shown in Ref.\cite{31,32}, such
energy-momentum redistribution resulting from scattering can substantially
affect superlattice properties and, in particular, the current-voltage
characteristics can change due to transverse heating. To overcome the
deficiency inherent in the single $\tau$-approximation, we develop a new
method based on the Boltzmann equation with a collision term encompassing
three distinct relaxation times. The three relaxation times include (a) a
time for redistribution of energy and momentum supplied by an electric field
to a given electron among its various degrees of freedom, (b) a time for
redistribution of energy and momentum among all electrons by inelastic
electron-electron interactions, and (c) a time for transfer of the excess
energy to the crystal lattice. In this, we employ a separation of the
relaxation processes into elastic, inelastic and electron-electron, which is
commonly recognized in the study of nonlinear properties of semiconductors
at high fields (see, for example, Ref.\cite{41,42}). The resulting balance
equations which we obtain can be solved analytically for systems having high
symmetry (two- and three-dimensional superlattices). However, the
qualitative results obtained here are also valid for bulk semiconductors
having a one-dimensional superlattice, i.e. for structures with minibands in
the growth direction and free electron motion in the lateral plane, which
can be studied experimentally currently. Moreover, at the present time,
three-dimensional cluster lattices are actually grown \cite{28} and
technological progress \cite{spr1,spr2,29} offers hope that two-dimensional
and three-dimensional superlattices will be fabricated in the near future
using quantum dots, relating directly to our studies here. Furthermore, a
simple three-dimensional Kronig-Penney model was proposed in order to
describe such quantum dots superlattices theoretically \cite{Laz}. On the
basis of the three relaxation time description, we show that electron
heating in the plane perpendicular to the current drastically changes the
self-induced transparency of the superlattice. In particular, it leads to an
additional modulation of the current amplitudes excited by an applied
biharmonic electric field with harmonic components polarized in orthogonal
directions. We obtain analytical results in the weak scattering
approximation $(\omega\tau >> 1)$ and extend the analysis numerically for
stronger scattering. }

{\normalsize This article is structured as follows. In Section II, starting
from a Boltzmann equation, we derive balance equations for average electron
velocity (current) and electron energy by means of the new collision term
accounting for momentum and energy redistribution among the various degrees
of freedom. On the basis of these balance equations the theory of
self-induced transparency and its mechanisms is developed in Section III for
one-, two- and three-dimensional superlattices in the presence of a
high-frequency (hf) harmonic field. In Section IV we analyze the amplitude
modulation of hf current by an orthogonal hf field of a different frequency.
The main results of this work and comparison to previous studies are
presented in Section V. }

{\normalsize \vspace*{1cm} }

{\large {\bf II. General Relations} }

{\normalsize \vspace*{1cm} }

{\normalsize There are several prominent sources of nonlinear electron
response in superlattices. Electron dynamics in narrow minibands features
phenomena manifested as Bloch oscillations, static electron localization and
the Wannier-Stark ladder \cite{4,5} in a static electric field; also, in a
harmonic electric field there are nonlinear oscillations with amplitudes
modulated by Bragg reflections and dynamic electron localization \cite{22,23}
with the collapse of the electron's minibands \cite{27}. In the case of
interminiband transitions, nonlinear phenomenology includes interminiband
tunneling in a static electric field, with or without photon-assistance \cite
{34}, also interminiband tunneling in a harmonic electric field and Rabi
oscillations \cite{35}. Of course, there are also nonlinear electron
response properties involved in the relaxation processes that redistribute
energy supplied by the electric field among the various degrees of freedom,
controlling the anisotropic heating of the electron gas \cite{31,32}, upon
which our attention is focused in this paper. }

{\normalsize We revisit the analysis of electron dynamics in a single
miniband from a different perspective than that of earlier studies \cite
{22,23,27}. In this, we will examine electron dynamics in two-dimensional $%
(\mu =2)$ and three- dimensional $(\mu =3)$ superlattices in the presence of
an oscillatory electric field having $\mu$ frequency components, 
\begin{equation}
{\bf E}(t) = \sum_{\alpha=1}^{\mu}{\bf x_{\alpha}}E_{\alpha}cos(\omega_{%
\alpha}t- \delta_{\alpha}),\hspace*{0.5cm} \mu=2,3,
\end{equation}
where ${\bf x_{\alpha}}$ are the unit lattice vectors of the crystal, $%
\delta_{\alpha}$ are the initial phases of the fields and the frequencies $%
\omega_{\alpha}$ are different in general. We take the electron energy
dispersion relation in the tight-binding approximation as: 
\begin{equation}
\varepsilon ({\bf k}) =
\sum_{\alpha=1}^{\mu}\varepsilon_{\alpha}(k_{\alpha}), \hspace*{0.5cm}
\varepsilon_{\alpha}(k_{\alpha}) = {\frac{\Delta}{2}} \left( 1-
cos(k_{\alpha}d)\right) ,
\end{equation}
where $\Delta$ is the miniband width, $\varepsilon_{\alpha}$ and $k_{\alpha}$
are the energy and wave number along the $\alpha$-axis, respectively. }

{\normalsize Under the influence of the electric field of Eq.(1), an
electron in a $\mu$-dimensional superlattice executes nonlinear oscillations
(with a different period in each direction) having the velocity given by (no
scattering) 
\begin{equation}
V_{\alpha}(k_{\alpha}^{(0)},t_0,t) = V_m \sum_{n=-
\infty}^{\infty}J_n(g_{\alpha})sin\left[ n(\omega_{\alpha}t-\delta_{\alpha})
+ k_{\alpha}^{(0)}d - g_{\alpha}sin(\omega_{\alpha}t_0)\right] ,
\end{equation}
where $V_m=\Delta d/2\hbar$ is the maximum electron velocity, $%
k_{\alpha}^{(0)}$ is the electron wave vector at initial time $t_0$, and $%
g_{\alpha}=\Omega^{(0)}_{\alpha}/\omega_{\alpha}=
eE_{\alpha}d/\hbar\omega_{\alpha}$. One can obtain a similar expression for
electron energy by integrating the relation $V_{\alpha}(k_{\alpha})=\hbar^{-
1}\partial\varepsilon_{\alpha}(k_{\alpha})/\partial k_{\alpha}$. }

{\normalsize It is clear from Eq.(3) that the velocity harmonic amplitudes
are oscillatory functions of the field amplitude. They sequentially vanish
at the zeros of the $J_n(g_{\alpha})$-functions independently of the initial
electron momentum. In particular, the time averaged values of electron
velocity and energy (zero$^{th}$ harmonics) are given by: 
\begin{equation}
\overline{V_{\alpha}}(k_{\alpha}^{(0)},t_0) = V_mJ_0(g_{\alpha})sin\left(
k_{\alpha}^{(0)}d-g_{\alpha}sin(\omega_{\alpha}t_0)\right),
\end{equation}
and 
\begin{equation}
\overline{\varepsilon_{\alpha}}(k_{\alpha}^{(0)},t_0) = {\frac{\Delta}{2}}%
\left[ 1 - J_0(g_{\alpha})cos\left( k_{\alpha}^{(0)}d-
g_{\alpha}sin(\omega_{\alpha}t_0)\right)\right] .
\end{equation}
At the specific values of $g_{\alpha}$ for which $J_0(g_{\alpha})=0$, we
have 
\begin{equation}
\overline{V_{\alpha}}(k_{\alpha}^{(0)},t_0) = 0, \hspace*{0.5cm} \overline{%
\varepsilon_{\alpha}}(k_{\alpha}^{(0)},t_0) = {\frac{\Delta}{2}},
\end{equation}
i.e. electron motion along the $\alpha$-axis has no net translation
independently of its initial momentum, and its average energy takes the
value at the middle of the one-dimensional miniband. This phenomenon is
known as {\it dynamic electron localization}. The electron has a discrete
energy spectrum due to the finiteness of the motion and, therefore, dynamic
electron localization corresponds to the collapse of its quasienergy
minibands \cite{27} [described by the pre-collapse relation 
\begin{equation}
\left. \tilde{\varepsilon}_{\alpha}(k_{\alpha}) = {\frac{\Delta}{2}}\left[ 1
- J_0(g_{\alpha})cos(k_{\alpha}d)\right] + n_{\alpha}\hbar\omega_{\alpha}, 
\hspace*{0.5cm} n_{\alpha} = 0, \pm 1, \pm 2, ... \right] .
\end{equation}
Depending on the values of $g_{\alpha}$, dynamic electron localization and
miniband collapse can be one-dimensional, two-dimensional, or
three-dimensional (in which case the localization and collapse are
complete). }

{\normalsize The dynamical peculiarities of superlattice electrons are
evident in nonlinear conduction. However, even a qualitative analysis
requires the use of a correct model of the relaxation processes, which may
be simplified to the specifications of a particular problem. In the present
work we endeavor to take account of the $\mu$-dimensionality of the electron
scattering processes and the separation of elastic and inelastic scattering.
The single $\tau$-approximation, useful for the one-dimensional model, is
not adequate for our purposes, as discussed above. A two relaxation time
model was proposed in Ref.\cite{39} as well as in Ref.\cite{16} by one of
the authors of the present work. However, this model is, in fact,
one-dimensional and, moreover, it produces the {\it illusion} of a
separation of elastic and inelastic electron scattering processes and
associated scattering times, which we now understand to be incorrect. In
some sense this model is even worse than usual single $\tau$-approximation
(but, unfortunately, is still in use (Ref.\cite{40})) because its
identification of the two relaxation times from experimental data is
erroneous. The balance equation method, developed in Ref.\cite{33}, would be
useful for our goals, if it would be generalized by replacing the isotropic
electron temperature by an anisotropic one (Ref.\cite{32}), but this
generalization presents a considerable challenge and has not been done yet.
The approach proposed in Ref.\cite{MS} takes into consideration the
interplay between different degrees of freedom and separates phonon and
impurity scatterings on a microscopic basis. However, this method is
essentially single-particle in nature and it is primarily applicable for a
superlattice with low carrier concentration when electron-electron
scattering does not play a significant role. To overcome these limitations,
we start from a three-dimensional model of a superlattice having the novel
phenomenological collision term proposed in Ref.\cite{32}. This collision
term describes scattering in terms of an improved three-channel electron
relaxation process. We employ the commonly understood \cite{41,42}
separation of the relaxation processes into elastic, inelastic and
electron-electron with characteristic times specified for each of the three
channels working in parallel. In the first channel (usually the fastest one)
an electron is subject to redistribution of the additional energy and
momentum supplied by the applied electric field among its degrees of freedom
by means of elastic scattering during a characteristic time $\tau_1$. The
kinetic energy of each electron is conserved during this scattering to
isoenergetic surfaces, but the {\it direction} of momentum is randomized
(with consequent reduction of the drift velocity). In the longer-lasting
second channel, the energy supplied by the external electric field is
redistributed among all electrons due to inelastic electron-electron
scattering, including Umklapp processes. As a result of the Umklapp
processes the Fermi distribution becomes {\it undrifted} \cite{42} and,
furthermore, the redistribution of energy establishes an effective electron
temperature $T_e$ during a time $\tau_{ee}$ by electron-electron scattering.
The total energy of all electrons is conserved during the redistribution
process, in spite of their momentum relaxation. Finally, in the third
channel, electrons transmit energy to the lattice over a time $%
\tau_{\varepsilon}$ and their distribution relaxes to a Fermi function at
the lattice temperature $T_0$. }

{\normalsize The Boltzmann equation with this model collision term has the
form: 
\begin{equation}
{\frac{\partial f({\bf k},t)}{\partial t}} + {\frac{e{\bf E}(t)}{\hbar}}{%
\frac{\partial f({\bf k},t)}{\partial k}} = -\left( {\frac{\partial f}{%
\partial t}}\right)_{coll} ,
\end{equation}
where 
\begin{equation}
\left( {\frac{\partial f}{\partial t}}\right)_{coll} = {\frac{f({\bf k}%
,t)-f_S(\varepsilon ,t)}{\tau_1}} + {\frac{f({\bf k},t)-f_0(\varepsilon ,T_e)%
}{\tau_{ee}}} + {\frac{f({\bf k},t)-f_0(\varepsilon ,T_0)}{\tau_{\varepsilon}%
}} ,
\end{equation}
with the isoenergetic distribution function, $f_S(\varepsilon ,t)$,
expressed as an integral average over the equipotential surface $%
S_{\varepsilon}$, 
\begin{equation}
f_S(\varepsilon ,t) = {\frac{\int_{S_{\varepsilon}}f({\bf k},t){\frac{dS}{%
\vert\nabla_k\varepsilon\vert}}}{\int_{S_{\varepsilon}}{\frac{dS}{%
\vert\nabla_k\varepsilon\vert}}}} ,
\end{equation}
and 
\begin{equation}
\langle\varepsilon\rangle = \langle\varepsilon\rangle_S =
\langle\varepsilon\rangle_e,
\end{equation}
where we use the notation $\varepsilon =\varepsilon ({\bf k})$ and $f({\bf k}%
,t)$ is the nonequilibrium distribution function driven by the electric
field. $f_0(\varepsilon ,T_e)$ is the equilibrium Fermi distribution at the
elevated electron temperature $T_e$ and $f_0(\varepsilon ,T_0)$ is the
equilibrium Fermi distribution at lattice temperature $T_0$. $%
\langle\varepsilon\rangle , \langle\varepsilon\rangle_S,
\langle\varepsilon\rangle_e$, and $\langle\varepsilon\rangle_0$ are the
energies averaged over the corresponding distribution functions. The
effective electron temperature $T_e$ is determined by Eq.(11). It is
important to note that the anisotropic function $f_S(\varepsilon ,t)$ plays
the role of the "isotropic" distribution function of Ref.\cite{41}, but not
the symmetric one, i.e. $f_S(\varepsilon ,t) \neq (f({\bf k},t)+ f(-{\bf k}%
,t))/2 $. A symmetric form of $f_S(\varepsilon ,t)$ would occur is there
were {\it no} energy and momentum redistribution among all degrees of
freedom and it would correspond to the one-dimensional model of a
superlattice used in Refs. \cite{16,39}. }

Let us summarize our generalizations and simplifications of the three
relaxation processes:

(1) As in Ref.\cite{41}, the quasielastic electron scattering leading to the
"isotropization" of the electron distribution function is considered to be
dominant. In our case, the electrons are distributed onto corresponding
(nonspherical!) isoenergetic surfaces and the anisotropic function $%
f_s(\varepsilon ,t)$ plays the role of the "isotropic" distribution function
of Ref.\cite{41}.

(2) The Brillouin minizones are narrow for superlattices, lending importance
to Umklapp processes and the establishment of an {\it undrifted} Fermi
distribution with an effective temperature $T_e$ \cite{42}. This underscores
the difference of our present description from that of Ref.\cite{33}.

(3) The dynamical development of the deviation of the electron distribution
function from the "isotropic" one is described by 
\begin{equation}
{\frac{\partial}{\partial t}}\left( f(k,t)-f_s(\varepsilon ,t)\right)_{st}=-{%
\frac{f(k,t)-f_s(\varepsilon ,t)}{\tau}} ,
\end{equation}
i.e. by the effective relaxation time $\tau$, which is, in general,
dependent not only on energy but also on the electron momentum direction.
This relation is the same as the one commonly used for the first term of the
distribution function expansion in Legendre polynomials \cite{41}. Thus, in
this respect, our description is the same as that of Ref.\cite{41} up to
this point. In accordance with our classification of the three relaxation
processes, we have 
\begin{equation}
{\frac{1}{\tau}} = {\frac{1}{\tau_1}}+{\frac{1}{\tau_{ee}}}+{\frac{1}{%
\tau_{\varepsilon}}}.
\end{equation}

(4) We describe the collision dynamics of the "isotropic" distribution
function $f_s(\varepsilon ,t)$ approximately by two relaxation times, $%
\tau_{ee}$ and $\tau_{\varepsilon}$, i.e. by the relation 
\begin{equation}
\left( {\frac{\partial f_s(\varepsilon ,t)}{\partial t}}\right)_{st} = - {%
\frac{f_s(\varepsilon ,t)-f_0(\varepsilon ,T_e)}{\tau_{ee}}} - {\frac{%
f_s(\varepsilon ,t)-f_0(\varepsilon ,T_0)}{\tau_{\varepsilon}}}.
\end{equation}
While this is not an exact equation for $f_s(\varepsilon ,t)$, it is
acceptable for a qualitative description. Moreover, our interest is not in
the details of the distribution functions, but only in the current and in
the average energy. Furthermore, it is easier to incorporate necessary
corrections directly in balance equations to be derived below.

The balance equations can be obtained from Eqs.(8-11). For sake of
simplicity we take the relaxation times $\tau_1$, $\tau_{ee}$, and $%
\tau_{\varepsilon}$ to be energy and momentum independent. Multiplying
Eq.(8) sequentially by $\partial\varepsilon_{\alpha}(k_{\alpha})/\partial
k_{\alpha}$ and by $\varepsilon_{\alpha}(k_{\alpha})$ and integrating over
first Brillouin minizone, we obtain the following equations for the current
components $\left( j_{\alpha}(t) =
ne\hbar^{-1}\left\langle\partial\varepsilon_{\alpha}(k_{\alpha})/\partial
k_{\alpha}\right\rangle\right)$ and the average energies $%
\langle\varepsilon_{\alpha}\rangle$: 
\begin{equation}
{\frac{\partial j_{\alpha}(t)}{\partial t}} - ne^2\left\langle
m_{\alpha}^{-1}(\varepsilon )\right\rangle E_{\alpha}(t) = -{\frac{%
j_{\alpha}(t)}{\tau_p}};
\end{equation}
\begin{equation}
{\frac{d}{dt}}\left\langle\varepsilon_{\alpha}\right\rangle - {\frac{1}{n}}%
E_{\alpha}(t)j_{\alpha}(t) = -{\frac{\left\langle\varepsilon_{\alpha}\right%
\rangle -\left\langle\varepsilon\right\rangle_S}{\tau_1}} -{\frac{%
\left\langle\varepsilon_{\alpha}\right\rangle
-\left\langle\varepsilon\right\rangle_e}{\tau_{ee}}} -{\frac{%
\left\langle\varepsilon_{\alpha}\right\rangle
-\left\langle\varepsilon\right\rangle_0}{\tau_{\varepsilon}}};
\end{equation}
and 
\begin{equation}
\sum_{\alpha =1}^{\mu}\left\langle\varepsilon_{\alpha}\right\rangle_S =
\sum_{\alpha =1}^{\mu}\left\langle\varepsilon_{\alpha}\right\rangle_e =
\langle\varepsilon\rangle ,
\end{equation}
where 
\begin{equation}
\left\langle m_{\alpha}^{-1}(\varepsilon )\right\rangle = \left\langle {%
\frac{1}{\hbar^2}}{\frac{\partial^2\varepsilon (k_{\alpha})}{\partial
k^2_{\alpha}}}\right\rangle ,
\end{equation}
$n$ is the three-dimensional electron density and $\tau_p^{-1} =
\tau_1^{-1}+\tau_{ee}^{-1}+\tau_{\varepsilon}^{-1}$ is the overall inverse
electron relaxation time.

The balance equations (Eqs.(15-17)) are valid both for homogeneous
semiconductors (without superlattice) and for semiconductor superlattices of
any dimensions. All terms in these equations have clear physical meaning and
allow some generalizations, for example, the replacement of scalar
relaxation times by a relaxation tensor (for symmetric structures we
consider it unnecessary). It should be noted that the relaxation times, $%
\tau_p$ and $\tau_{\varepsilon}$, can be taken from independent calculations
using the actual scattering mechanisms. Such a calculation for
one-dimensional GaAs-based superlattices was done in Ref.\cite{Lei}, where,
in particular, it was shown that the relaxation times, $\tau_p$ and $%
\tau_{\varepsilon}$, can be taken to be independent of energy, if the
miniband width is less than optical phonon energy.

In general, the set of equations Eqs.(15-17) is not closed because of
coupling to higher order moments of the distribution function $f({\bf k},t)$%
. However, there are simplifications for one-, two- and three-dimensional
superlattices with a sinusoidal dispersion law, Eq.(2). In these cases,
symmetry dictates that 
\begin{equation}
\left\langle\varepsilon_{\alpha}\right\rangle_S =
\left\langle\varepsilon_{\alpha}\right\rangle_e = {\frac{1}{\mu}}%
\left\langle\varepsilon\right\rangle .
\end{equation}
Furthermore, the sinusoidal dispersion law provides the linear
proportionality between the effective electron mass and its energy: 
\begin{equation}
\left\langle m_{\alpha}^{-1}(\varepsilon )\right\rangle = \left( {\frac{%
\Delta}{2}}- \left\langle\varepsilon_{\alpha}\right\rangle \right){\frac{d^2%
}{\hbar^2}} .
\end{equation}
Accordingly, Eqs.(15,16), taken jointly with Eqs.(19,20), form a closed set
of equations. It is convenient to write this set in complex form,
introducing a dimensionless complex $\mu$-component "vector" with components
defined by 
\begin{equation}
\Phi_{\alpha}(t) = {\frac{\Delta /2-\langle\varepsilon_{\alpha}\rangle}{%
\Delta /2- \langle\varepsilon_{\alpha}\rangle_0}} - i{\frac{j_{\alpha}(t)}{%
j_{0\alpha}}} ,
\end{equation}
where $j_{0\alpha}=(end/\hbar )(\Delta
/2-\langle\varepsilon_{\alpha}\rangle_0)$. The balance equation for $%
\Phi_{\alpha}(t)$, equivalent to Eqs.(15,16,19,20), is given by 
\begin{equation}
{\frac{d\Phi_{\alpha}(t)}{dt}} +
(\tau_p^{-1}+i\Omega_{\alpha}(t))\Phi_{\alpha}(t) = \tau_{\varepsilon}^{-1}
+ {\frac{1}{\mu}}(\tau_p^{-1}-\tau_{\varepsilon}^{-
1})\sum_{\beta=1}^{\mu}Re\Phi_{\beta}(t),
\end{equation}
where $\Omega_{\alpha}(t)=edE_{\alpha}(t)/\hbar$. The last term on the right
side of Eq.(22) describes the redistribution of electron energy and momentum
among all degrees of freedom. This feature is absent in the single
relaxation time description and is of crucial importance for our present
considerations. For a one-dimensional superlattice, ($\mu =1$), Eqs.(22) are
identical to the balance equations obtained in Ref.\cite{16}. It should be
noted that the only significant feature of the distribution functions, $%
f_S(\varepsilon ,t)$ and $f_0(\varepsilon ,T_e)$, is that given by Eq.(19)
in regard to the derivation of Eq.(22) for two- and three-dimensional
superlattices, because of the high symmetry of the electron dispersion
relation. The specific forms of $f_S(\varepsilon ,t)$ and $f_0(\varepsilon
,T_e)$, beyond Eq.(19), are not pertinent. A further consequence of this
symmetry is that the inverse relaxation times $\tau_1^{-1}$ and $%
\tau_{ee}^{-1}$ are involved in Eq.(22) only in the form of their sum. For
bulk semiconductors having a one-dimensional superlattice with free motion
in the lateral plane, one obtains an integro-differential equation (allowing
only numerical solution) instead of Eq.(22). Such an integro-differential
equation was analyzed for the case of a static electric field in Ref.\cite
{32} in detail.

{\normalsize It is easily shown that the complex ''vector'' with components $%
\Phi _{\alpha }(t)$ introduced above is constituted by the first
Fourier-components of distribution function $f({\bf k},t)$. Using
periodicity in ${\bf k}$-space, this distribution function may be expanded
in a $\mu $-dimensional Fourier series: 
\begin{equation}
f({\bf k},t)=\sum_{\nu _{\alpha }}F_{{\bf {\nu }}}exp\{i{\bf \nu \cdot k}d%
{\bf \}}\Phi _{{\nu }}(t){\bf ,}
\end{equation}
with Fourier coefficients given by 
\begin{equation}
F_{{\bf {\nu }}}=\left( {\frac{d}{2\pi }}\right) ^{\mu }\int_{BZ}d^{\mu
}kf_{0}({\bf k})exp\{-i{\bf \nu \cdot k}d{\bf \},}
\end{equation}
where ${\bf \nu }\rightarrow (\nu _{1}{\bf )}$ for one-dimensional,} ${\bf %
\nu }\rightarrow (\nu _{1},\nu _{2})${\normalsize \ for two-dimensional, }$%
{\bf \nu }\rightarrow (\nu _{1},\nu _{2},\nu _{3})${\normalsize \ for
three-dimensional superlattices, and the integration is taken over the first
Brillouin zone (BZ). Only the first momentum harmonics (in any direction) of
the distribution function 
\begin{equation}
\Phi _{\alpha }(t)={\frac{\int d^{\mu }kf({\bf k},t)exp\{-ik_{\alpha }d\}}{%
\int d^{\mu }kf_{0}({\bf k})exp\{-ik_{\alpha }d\}}}
\end{equation}
contribute to the current density ${\bf j}(t)$ and the electron energy $%
\langle \varepsilon _{\alpha }\rangle $ for the miniband case of Eq.(2). One
can see that they are the same as those given by Eq.(21). }

{\normalsize For arbitrary time-dependence of the electric field ${\bf E}(t)$%
, it is useful to write $\Phi _{\alpha }(t)$ in a form that is convenient
for the representation of Bloch oscillations, as 
\begin{equation}
\Phi _{\alpha }(t)=a_{\alpha }(t)\Psi _{\alpha }(t),
\end{equation}
where 
\begin{equation}
\Psi _{\alpha }(t)=exp\left\{ -i\int_{0}^{t}\Omega _{\alpha
}(t_{1})dt_{1}\right\}
\end{equation}
is a solution of the homogeneous counterpart of Eq.(22) associated with
Bloch oscillations in the absence of scattering terms. This function
describes the dynamic modulation of the electron distribution function by
the applied electric field without scattering. In particular, for a simple
harmonic field, $E_{\alpha }(t)=E_{\alpha }^{(0)}cos\omega t$, we have 
\begin{equation}
\Psi _{\alpha }(t)=exp\{-ig_{\alpha }sin\omega t\}=\sum_{\nu =-\infty
}^{\infty }J_{\nu }(g_{\alpha })e^{-i\nu \omega t},
\end{equation}
where $g_{\alpha }=\Omega _{\alpha }^{(0)}/\omega $ is the projection of the
dimensionless field amplitude onto the $x_{\alpha }$-axis. The functions $%
a_{\alpha }(t)$ represent dissipative processes, describing changes in the
amplitude of $\Phi _{\alpha }(t)$ (deviation from the Bloch oscillation
solution) due to scattering. They obey the following equations, based on
Eq.(22), 
\begin{equation}
\dot{a}_{\alpha }(t)+\tau _{p}^{-1}a_{\alpha }(t)=\tau _{\varepsilon
}^{-1}\Psi _{\alpha }^{\ast }(t)+{\frac{1}{\mu }}\left( \tau _{p}^{-1}-\tau
_{\varepsilon }^{-1}\right) Re\left( \sum_{\beta =1}^{\mu }a_{\beta }\Psi
_{\beta }\right) \Psi _{\alpha }^{\ast }(t).
\end{equation}
In the absence of scattering, $a_{\alpha }(t)\equiv 1$. The transfer from
the description in terms of the functions $\Phi _{\alpha }(t)$ to a
description in terms of functions $a_{\alpha }(t)$ corresponds to a
transformation to a new system of coordinates, $K_{0}$, oscillating in
momentum space together with the unscattered electron. In the system $K_{0}$%
, each electron is at a fixed point ${\bf k_{0}}$. Only scattering changes
the distribution of these points. In the case of rare collisions $(\omega
\tau >>1)$, such changes are small during the period of the applied field,
but they can accumulate and become important over the time of a few
collisions. Otherwise, the equilibrium distribution functions (to which the
electrons relax) in the system $K_{0}$ are modulated by the field and become
rapid functions of time, as 
\begin{equation}
f_{0}(\varepsilon ,T_{0})\rightarrow f_{0}(k_{3})=f_{0}\left( {\bf k_{0}}-{%
\frac{1}{d}}\int_{0}^{t}{\bf \Omega }(t_{1})dt_{1}\right) .
\end{equation}
This feature is embodied in the structure of Eq.(29): the ''scattering-out''
term (second term on the left side) has the usual relaxation form with the
overall inverse relaxation time due to all scattering mechanisms, and the
''scattering-in'' term (first term on the right side) is the dynamically
modulated equilibrium distribution function with inverse relaxation time $%
\tau _{\varepsilon }^{-1}$. The last term on the right side of Eq.(29),
describing the redistribution of energy and momentum over the various
degrees of freedom, is modulated by the electric field twofold: once in
connection with the transformation to the system $K_{0}$, and, secondly,
because the corresponding equilibrium distribution functions are determined
by the average normalized electron energy (Eqs.(21),(26)) which, in turn,
depends on the field and time. This average energy is involved in the last
term through the relation 
\[
Re\left( \sum_{\beta =1}^{\mu }a_{\beta }\Psi _{\beta }\right) =Re\left(
\sum_{\beta =1}^{\mu }\Phi _{\beta }(t)\right) =\sum_{\beta =1}^{\mu }{\frac{%
\Delta /2-\langle \varepsilon _{\beta }\rangle }{\Delta /2-\langle
\varepsilon _{\beta }\rangle _{0}}}. 
\]
\vspace*{1cm} }

{\large {\bf III. Self-induced transparency} }

{\normalsize \vspace*{1cm} }

{\normalsize In this section, we analyze superlattice response to a high
frequency $(\omega\tau_p>>1)$ monochromatic field directed along the $x_1$%
-axis. Considering $a_{\alpha}$ to be slowly varying functions of time, we
average Eq.(29) over an interval $\Delta t$ given by $2\pi /\omega << \Delta
t << \tau_p$, obtaining an algebraic system of equations, for which the
stationary solution is 
\begin{equation}
a_1 = {\frac{\lambda\mu^2J_0(g_1)}{\mu\left[ 1+(\mu -1)\lambda\right]\cdot
B(g_1) - (\mu -1)(1-\lambda)^2J_0^2(g_1)}},
\end{equation}
and 
\begin{equation}
a_{\alpha} = \lambda {\frac{\mu B(g_1) + (1-\lambda) J_0^2(g_1)}{\left[
1+(\mu - 1)\lambda\right]\cdot B(g_1) - (\mu -1)(1-\lambda)^2J_0^2(g_1)}}, 
\hspace*{0.5cm} \alpha\neq 1,
\end{equation}
where $\lambda = \tau_p/\tau_{\varepsilon}$ and 
\begin{equation}
B(g) = 1 - {\frac{1-\lambda}{2\mu}}\left[ 1 + J_0(2g)\right] .
\end{equation}
In the derivation of Eqs.(31),(32), we used the relations 
\begin{equation}
\overline{\Psi_{\alpha}(t)} = J_0(g_{\alpha}), \hspace*{1cm} \overline{%
(Re(\Psi_{\alpha}(t))^2} = {\frac{1}{2}}\left[ 1 + J_0(2g_{\alpha})\right] ,
\end{equation}
where the overhead bar indicates averaging over the period of the impressed
electric field. }

{\normalsize According to Eqs. (21),(26),(28),(31) and (32), the current,
energy dissipation rate, $Q$, and ratio between the transverse and
longitudinal electron heating, $\delta$, are given by 
\begin{equation}
\tilde{j_1}\equiv j_1/j_{01} = a_1sin(g_1 sin(\omega t)) + O\left( {\frac{1}{%
\omega\tau}}\right) ;
\end{equation}
\begin{equation}
Q\equiv n{\frac{\overline{\langle\varepsilon\rangle} -
\langle\varepsilon\rangle_0 }{\tau_{\varepsilon}}} = Q_0\left( 1 -
a_1J_0(g_1)\right) , \hspace*{0.5cm} Q_0 = {\frac{\mu\left( \Delta /2 -
\langle\varepsilon_{\alpha}\rangle_0\right) n}{\tau_{\varepsilon} + (\mu
-1)\tau_p}};
\end{equation}
and 
\begin{equation}
\delta\equiv {\frac{(\mu -1)\left( \overline{\langle\varepsilon_{\alpha}%
\rangle} - \langle\varepsilon_{\alpha}\rangle_0\right)}{\overline{%
\langle\varepsilon_1\rangle} - \langle\varepsilon_1\rangle_0}} = {\frac{%
(1-\lambda )(\mu -1)}{1 + \lambda(\mu - 1)}}.
\end{equation}
}

{\normalsize Figure 1 depicts the function $a_1$ involved in the expression
for the current, and also shows $Q/Q_0$, as functions of $g$ for a
three-dimensional superlattice $(\mu =3)$ with $\lambda = 1, 0.1, 0.01$. In
the case $\tau_p = \tau_{\varepsilon} (\lambda =1)$, we obtain $a_1 =
J_0(g_1)$ and expressions (35) and (36) coincide with those obtained for the
one-dimensional model of a superlattice \cite{16}. With increasing $\tau_p$
while holding $\tau_{\varepsilon}$ fixed, the current in the superlattice
increases, the dissipation decreases, and the amplitude of their modulation
increases. As in the case of the usual single $\tau$-approximation \cite
{23,16,24}, the three-dimensional superlattice current vanishes at fields
such that $J_0(g_1)=0$ (with an accuracy of $(\omega\tau_p)^{-1}$), where
dissipation is maximal. When elastic collisions dominate over inelastic
scattering $(\tau_p<<\tau_{\varepsilon})$ there is a rapid redistribution of
energy among the electron degrees of freedom and, consequently, the current
decreases sharply for $g_1\geq 1$. This is caused by a strong expansion of
the distribution function in momentum space, due to both longitudinal and
transverse heating. }

{\normalsize To examine the peculiarities of self-induced transparency in
the case of a two-dimensional superlattice, we consider an electric field of
the form, 
\begin{equation}
{\bf E} = (E_1{\bf x_1} + E_2{\bf x_2})cos(\omega t).
\end{equation}
In a manner similar to that of the foregoing analysis, we obtain the
components of the dissipative function as, 
\begin{equation}
a_{1,2} = {\frac{4\lambda\left( J_0(g_{1,2})B(g_{1,2})+(1-\lambda
)J_0(g_{1,2})(J_0(g_1+g_2)+J_0(g_1-g_2))\right)}{16B(g_1)B(g_2)-(1-\lambda
)^2(J_0(g_1+g_2)+J_0(g_1-g_2))^2}} .
\end{equation}
To obtain Eq.(39), we used the relations (34) and the expression 
\begin{equation}
\overline{Re\Psi_1(t)Re\Psi_2(t)} = {\frac{1}{2}}\left(
J_0(g_1+g_2)+J_0(g_1- g_2)\right) .
\end{equation}
The corresponding current and energy dissipation rate are given by 
\begin{equation}
j_{1,2}(t) = a_{1,2}sin(g_{1,2}sin\omega t) ,
\end{equation}
and 
\begin{equation}
Q = {\frac{n\left(\Delta - \langle\varepsilon\rangle_0\right)}{%
\tau_{\varepsilon}}}\left( 1 - {\frac{1}{2}}(a_1J_0(g_1) +
a_2J_0(g_2))\right) .
\end{equation}
}

{\normalsize It is clear from Eqs.(39),(41) that the current along each
superlattice axis depends on all field projections, in contrast to the
results of Ref.\cite{44}. In general, the current of each excited harmonic
is not parallel or antiparallel to the resultant electric field and has its
own elliptic polarization. There are exceptions if the electric field is in
the lattice directions $\overline{[10]}$ or $\overline{[11]}$, in which case
the field and excited current are parallel. }

{\normalsize Furthermore, it is evident from Eqs.(39),(41), that both
dynamic localization and self-induced transparency can be either
one-dimensional or complete for two-dimensional superlattices. In Figure 2
we show the locii of one-dimensional self-induced transparencies occuring in
directions ${\bf x_1}$ (horizontal curves) and ${\bf x_2}$ (vertical
curves), respectively, as functions of $g_1$ and $g_2$. At fields
corresponding to these curves, the current components vanish ($j_1(t)\approx
0$ and $j_2(t)\approx 0$, respectively). These dependencies exhibit
oscillations around the line $J_0(g_{1,2}) = 0$ (see Figure 2 Insert), which
is the condition for dynamic localization to occur. The average energies and
the dissipation rate for these values of the fields are given by 
\begin{equation}
\overline{\langle\varepsilon_{1(2)}\rangle} = {\frac{\Delta}{2}}, \hspace*%
{0.5cm} \overline{\langle\varepsilon_{2(1)}\rangle} = {\frac{\Delta}{2}} -
\left( {\frac{\Delta}{2}} - {\langle\varepsilon_{2(1)}\rangle}_0\right) {%
\frac{4\lambda J_0^2(g_{2(1)})}{4 + (1-\lambda )\left(
1+J_0(2g_{2(1)})\right)}},
\end{equation}
and 
\begin{equation}
Q = {\frac{n\left( \Delta - {\langle\varepsilon\rangle}_0\right)}{%
\tau_{\varepsilon}}}\left( 1 - {\frac{2\lambda J_0^2(g_{2(1)})}{4 +
(1-\lambda )\left( 1+J_0(2g_{2(1)})\right)}}\right) ,
\end{equation}
where the first of the subscripted indices is related to the horizontal
curves and the second subscripted index (in parentheses) is related to the
vertical curves. At points of intersection, determined by the condition $%
J_0(g_1)=J_0(g_2)=0$, there is complete self-induced transparency. In such
cases, the average electron energies and the dissipation rate are maximal
and are given by 
\begin{equation}
\overline{\langle\varepsilon_1\rangle} = \overline{\langle\varepsilon_2%
\rangle} = {\frac{\Delta}{2}} ,
\end{equation}
and 
\begin{equation}
Q = {\frac{n\left( \Delta - {\langle\varepsilon\rangle}_0\right)}{%
\tau_{\varepsilon}}}.
\end{equation}
The dissipation rate in Eq.(46) is larger than that in Eq.(44), and is also
larger than the maximum dissipation rate for the fields oriented strictly
along the crystal axes, the latter being given by Eq.(36) as 
\begin{equation}
Q = {\frac{n\left( \Delta - {\langle\varepsilon\rangle}_0\right)}{%
\tau_{\varepsilon} + \tau_p}}.
\end{equation}
}

{\normalsize It is apparent that, within the three-relaxation-time
description presently under consideration, complete self-induced
transparency and dynamic localization occur at the same fields, whereas
one-dimensional self-induced transparency and dynamic localization arise at
different fields. In two-dimensional superlattices complete dynamic
localization and self-induced transparency occur at the discrete amplitude
values and applied electric field directions determined by the relations 
\begin{equation}
E_{m,n} = {\frac{\hbar\omega}{ed}}\sqrt{\xi_m^2+\xi_n^2}, \hspace*{0.5cm}
\varphi_{m,n} = arctan(\xi_m^2/\xi_n^2),
\end{equation}
where $\varphi_{m,n}$ is the angle of field orientation with respect to the
superlattice crystal axis and $\xi_m$ is the $m^{th}$-order root of the zero$%
^{th}$-order Bessel function. This can be generalized easily to the
three-dimensional case. }

{\normalsize To develop a physical understanding of one-dimensional and
complete self-induced transparencies, we examine the evolution of the
electron distribution in the $K_0$-system. One can see from Eq.(29) and its
following discussion that, under the influence of the field and scattering,
the number of electrons entering the current components, $Im\Phi_{1,2}$
(including redistribution among them), averaged over the field period, $%
P_{1,2}$, is given by 
\begin{equation}
P_{1,2} = \tau^{-1}_{\varepsilon}\overline{\Psi_{1,2}(t)} + {\frac{1}{\mu}}%
\left(\tau^{-1}_p - \tau^{-1}_{\varepsilon}\right)\left( a_{1,2}\overline{%
\left( Re\Psi_{1,2}(t)\right)^2} + a_{2,1} \overline{Re\Psi_1(t) Re\Psi_2(t)}
\right) .
\end{equation}
If the dynamic modulation of the equilibrium distribution function is such
that $\overline{\Psi_1(t)} = \overline{\Psi_2(t)} = 0$, then the average
number of electrons entering the current components vanishes (there is only
electron redistribution among the components), and, therefore, the current
components are eventually completely eliminated from the nonequlibrium
distribution function by "scattering-out" over a time of order $\tau_p$. It
should be noted that for $\overline{\Psi_1(t)} = \overline{\Psi_2(t)} = 0$,
the set of equations (29) averaged over the field period becomes homogeneous
and its steady-state solution is zero. In this case electron heating is
maximal due to complete dynamic localization. Thus, after a time of order $%
\tau_p$, the superlattice becomes transparent, i.e. behaves like a
dielectric having the permittivity of the crystal lattice and relatively
small, but resonant, absorption. This is to say that we have complete
self-induced transparency. It should be emphasized that, at arbitrary
fields, the absorption rate stabilizes after time $\tau_{\varepsilon}$, i.e.
later than the vanishing of the current. }

{\normalsize If dynamic localization takes place in only one of the two
crystal directions, for example, in $\overline{[10]}$, i.e. $\overline{%
Re\Psi_1(t)} = J_0(g_1) = 0$, but $\overline{Re\Psi_2(t)} = J_0(g_2) \neq 0$%
, then the electrons only enter the current component $\Phi_2(t)$ due to
dynamic modulation of the equilibrium distribution function. However,
because of the redistribution of energy and momentum among the degrees of
freedom, electrons also flow into $\Phi_1(t)$. Therefore, even at $J_0(g_1)
= 0$ both $P_1$ and $j_1$ are nonzero. The current $j_1$ vanishes only if
the components of the flow $P_1$ caused (a) by direct dynamical modulation
of the distribution function, and (b) by redistribution via scattering,
compensate each other in this direction. However, this occurs at $J_0(g_1)
\neq 0$, i.e. when dynamic localization in this direction is absent, as
reflected in Eq.(32) and Figure 2. Furthermore, it should be noted that the
time-averaged one-dimensional energy is $\Delta /2$ for each electron when
dynamic localization occurs, but this is valid only for both time and
ensemble averaged energy in the case of one-dimensional self-induced
transparency. }

{\normalsize One can easily see that if the relaxation times $\tau_p$ and $%
\tau_{\varepsilon}$ are energy dependent ({\it not} constant), then the
first harmonics of the distribution (23) become coupled to higher harmonics.
In this case self-induced transparency does not occur even when there is
complete dynamic localization. This was demonstrated in Ref.\cite{23} for a
one-dimensional superlattice. However, as mentioned above, for an
appropriate set of superlattice parameters, these relaxation times can be
considered energy-independent \cite{Lei} and, therefore, even the
quantitative results discussed above have a wide range of validity. }

{\normalsize It should be emphasized that, although both self-induced
transparency and dynamic localization occur in a superlattice due to the
narrowness of its Brillouin minizones, the physical origins of these effects
are completely different. Dynamic localization arises when the zero$^{th}$
harmonic of nonlinear electron oscillations, modulated by Bragg reflections,
vanishes. In contrast to dynamic localization, self-induced transparency is
a result of the joint action of Bragg reflections of miniband electrons and
collisions creating strongly modulated electron distributions in which the
first harmonics are absent for discrete values of the electric field
amplitudes (in the limit $\tau_p\rightarrow\infty$). As a result, dynamic
localization appears immediately after turn-on of electric field, and
self-induced transparency occurs only after a time of order of $\tau_p$. The
conditions for self-induced transparency depend on the scattering mechanisms
and, in general, it takes place even without dynamic localization. It can be
shown that states of self-induced transparency are not stable with respect
to the generation of static and hf fields (having frequencies not equal to $%
\omega_1$) \cite{21} and, therefore, experimental studies should be
performed at low electron concentrations and with the use of pulsed electric
fields. }

{\normalsize \vspace*{1cm} }

{\large {\bf IV. Current modulation by orthogonal fields} }

{\normalsize \vspace*{1cm} }

{\normalsize In this section we examine superlattice behavior in the
presence of a high frequency biharmonic electric field given by 
\begin{equation}
{\bf E} = E_1{\bf x_1}cos(\omega_1t-\delta_1) + E_2{\bf x_2}cos(\omega_2t-
\delta_2).
\end{equation}
In this case the field components have different frequencies and are
directed along different crystal axes taken to be orthogonal (the case of
parallel fields was analyzed in Ref.\cite{45} in a single $\tau$%
-approximation). We are interested to explore the occurrence of current
amplitude modulation by a high frequency electric field orthogonal to the
current direction. This is determined by the redistribution of energy and
momentum among the various degrees of freedom. We assume that the
frequencies of the most important electric field harmonics are well
separated, i.e. 
\begin{equation}
\vert n_1\omega_1 - n_2\omega_2\vert\tau_p >> 1,\hspace*{0.5cm}n_{1,2} =
1,2,...\hspace*{0.1cm}.
\end{equation}
Using Eq.(34) and the expression 
\begin{equation}
\overline{Re\Psi_{\alpha}(t) Re\Psi_{\beta}(t)} =
J_0(g_{\alpha})J_0(g_{\beta}), \hspace*{0.5cm} \alpha\neq\beta ,
\end{equation}
we obtain the following relations for current components and averaged
energies: 
\begin{equation}
j_{1,2} = a_{1,2}(g_1,g_2)sin(g_{1,2}sin(\omega_{1,2}t-\delta_{1,2})),
\end{equation}
and 
\begin{equation}
\overline{\langle\varepsilon_{1,2}\rangle}-\langle\varepsilon_{1,2}\rangle_0
= \left( {\frac{\Delta}{2}}-\langle\varepsilon_{1,2}\rangle_0\right)\left(1-
a_{1,2}(g_1,g_2) J_0(g_{1,2})\right) ,
\end{equation}
where, in the case of a two-dimensional superlattice, 
\begin{equation}
a_{1,2}(g_1,g_2) = {\frac{4\lambda J_0(g_{1,2})\left( 2B(g_{2,1})+(1-\lambda
)J_0^2(g_{2,1})\right)}{4B(g_1)B(g_2)-(1-\lambda )^2 J_0^2(g_1)J_0^2(g_2)}},
\end{equation}
whereas, for a three-dimensional superlattice we have 
\begin{eqnarray}
&&a_{1,2}(g_1,g_2) =  \nonumber \\
&&{\frac{9\lambda J_0(g_{1,2})\left( 3B(g_{2,1})+(1- \lambda
)J_0^2(g_{2,1})\right)}{9(2+\lambda )B(g_1)B(g_2)-(1-\lambda )^2 \left(
3J_0^2(g_1)B(g_2)+3J_0^2(g_2)B(g_1)+(4-\lambda ) J_0^2(g_1)J_0^2(g_2)\right)}%
}
\end{eqnarray}
}

{\normalsize In Figure 3a we exhibit the functions $\tilde{a}%
^{-1}_1(0.05,g_2)\equiv a_1(g_1=0.05,g_2=0)/a_1(g_1=0.05,g_2)$ and $\tilde{a}%
^{-1}_1(g_1,2.405)\equiv a_1(g_1,g_2=0)/a_1(g_1,g_2=2.405)$ for $\lambda=0.1$%
. The former function describes the modulation due to the electric field $E_2
$ (orthogonal to $E_1$) of current driven by a weak electric field $E_1$;
whereas the latter function describes the same current modulation in the
presence of an arbitrary field $E_1$ under conditions of dynamic
localization in the direction ${\bf x_2} (J_0(g_2)=0)$. The functions $%
a_1(g_1,g_2=0)$ and $a_1(g_1,g_2=2.405)$, corresponding to the curve $\tilde{%
a}^{-1}_1(g_1,2.405)$ of Figure 3a, are shown in Figure 3b. }

{\normalsize One can see from Eqs.(55),(56) and Figures 3a,b, that it is
impossible for the orthogonal field to cause a complete vanishing of
polarization, and, therefore, {\it induced} transparency does not occur.
However, the modulation of polarization by the orthogonal field can be
significant, especially for small $\lambda$. The maximum decrease of
polarization comes about with the occurrence of one-dimensional dynamic
electron localization in the transverse direction $(J_0(g_2)=0)$. In
particular, for $E_1\rightarrow 0$, it decreases by the factor $(1+(\mu
-1)\lambda )/\mu\lambda$. The reason for this is that, under condition of
dynamic localization, electron heating is maximal and the distribution
function widens in all directions in momentum space due to energy and
momentum redistribution among all degrees of freedom, which always leads to
decreased current. }

{\normalsize As in the case of self-induced transparency, for specific
ratios of field amplitudes and frequencies determined by the conditions $%
J_0(g_1)=J_0(g_2)=0$, complete induced superlattice transparency takes
place. In this case a two- dimensional superlattice is transparent to an
arbitrarily polarized third weak signal with frequency $\omega_3$ (well
separated from the frequencies $n_1\omega_1\pm n_2\omega_2$, where $n_1$ and 
$n_2$ are integers).}

Similar to self-induced transparency, current modulation by an orthogonal
field and induced transparency occur (vanish) in a time of order of $\tau_p$
and they become stationary in a time of order of $\tau_{\varepsilon}$ after
turn-on (turn-off) of the electric field, i.e they are displaced in time
from dynamic localization.

{\normalsize To establish the frequency limitations of the phenomena we have
explored, we carried out a numerical analysis of Eq.(22) with finite values
of $\tau_p$ and $\tau_{\varepsilon}$. The results are shown in Fig.4, where
the dashed lines represent the above-described analytical calculations for
the functions $\tilde{a}^{-1}_1(0.05,g_2)$ (upper curve) and $\tilde{a}%
^{-1}_1(g_1,2.405)$ (lower curve), respectively. The solid lines represent
our numerical determinations of the amplitudes of the first current
harmonics (thick curves) and the maximal currents (thin curves) at $%
\omega\tau_p=1$ (maintaining $\omega\tau_{\varepsilon}>>1$). One can see
that the dependencies presented in Figure 3(a,b) change only slightly with
decreasing $\tau_p$. \vspace*{1cm} }

{\large {\bf V. Summary} }

{\normalsize \vspace*{1cm} }

{\normalsize In summary, we have applied the Boltzmann equation with an
improved three-relaxation-time collision term to the analysis of
self-induced and induced transparencies in semiconductor superlattices. The
three relaxation times include (a) a time for redistribution of energy and
momentum supplied by an electric field to a given electron among its various
degrees of freedom, (b) a time for redistribution of energy and momentum
among all electrons by inelastic electron-electron interactions, and (c) a
time for transfer of the excess energy to the crystal lattice. We have
performed analytical calculations for systems having high symmetry (for
one-, two- and three-dimensional superlattices). However, the results
obtained here are valid qualitatively for bulk semiconductors with a
one-dimensional superlattice, which are currently available for
experimentation. Furthermore, we have shown that self-induced transparency
and dynamic localization are different phenomena with different physical
origins, displaced in time from each other, and, in general, they arise at
different electric fields. Moreover, we have found that the redistribution
of energy and momentum among the various degrees of freedom is of crucial
importance in two-dimensional and three-dimensional superlattice transport
and optical properties. Transverse electron heating drastically changes the
conditions for self-induced transparency, and this effect facilitates
current modulation by an applied perpendicular high-frequency electric
field. }

\vspace*{1cm}

{\large {\bf Acknowledgements} }

\vspace*{1cm}

The work of Yu.A.R. and J.Yu.R. is supported by the Russian Foundation for
Basis Research (Grant No. 01-02-16446) and Ministry of Industry, Science and Technology of Russian Federation, L.G.M. and N.J.M.H. gratefully
acknowledge support from the Department of Defense, DAAD 19-01-1-0592.

{\normalsize \newpage }

{\normalsize Captions to the Figures: \vspace*{2cm} }

{\normalsize Figure 1. Normalized (a) current and (b) energy dissipation
rate as functions of $g=eEd/\hbar\omega$. }

{\normalsize Figure 2. Locii for one-dimensional self-induced transparencies
in directions ${\bf x_1}$ (horizontal curves) and ${\bf x_2}$ (vertical
curves). The Inset exhibits a magnification of the locus for one-dimensional
self-induced transparency in direction ${\bf x_1}$ and the locus for
one-dimensional dynamic localization in this direction (horizontal line). }

{\normalsize Figure 3. (a) Current modulation by orthogonal fields as
described by $\tilde{a}^{-1}_1(0.05,g_2)$ and $\tilde{a}^{-1}_1(g_1,2.405)$
[defined in text]; (b) The functions $a_1(g_1,g_2=0)$ and $%
a_1(g_1,g_2=2.405) $ employed in the determination of $\tilde{a}%
^{-1}_1(g_1,2.405)$ of Fig.3a. }

{\normalsize Figure 4. Amplitude of $j/j_0$ as a function of field ($%
g_{1(2)} $). }


\begin{references}
\bibitem{1}  L.V. Keldysh, Sov. Phys. Solid State {\bf 4}, 1658 (1962).

\bibitem{2}  {\normalsize L. Esaki and R. Tsu, IBM J. Res. Dev. {\bf 14}, 61
(1970). }

\bibitem{3}  M.I. Ovsyannikov, Yu.A. Romanov, V.N. Shabanov, and
P.G.Loginova, Sov. Phys. Semicond. {\bf 4}, 1919 (1970); Yu.A. Romanov, Sov.
Phys. Semicond. {\bf 5}, 1256 (1971).

\bibitem{4}  F. Bloch, Z. Phys. {\bf 52}, 555 (1928); C. Zener, Proc. R.
Soc. London Ser. A {\bf 145}, 523 (1934).

\bibitem{5}  G.H. Wannier, Phys. Rev. {\bf 117}, 432 (1950); Rev. Mod. Phys. 
{\bf 34}, 645 (1962).

\bibitem{22}  {\normalsize D.H. Dunlap and V.M. Kenkre. Phys. Rev. B {\bf 34}%
, 3625 (1986); Phys.Lett.A{\bf 127}, 438 (1988). }

\bibitem{23}  {\normalsize Yu.A. Romanov and Yu.Yu. Romanova, Phys. Solid
St. {\bf 43}, 539 (2001). }

\bibitem{16}  {\normalsize A.A. Ignatov and Yu. A. Romanov, Phys. St. Solidi
B {\bf 73}, 327 (1976). }

\bibitem{24}  {\normalsize A.A. Ignatov and Yu. A. Romanov, Sov. Phys. Solid
State {\bf 17}, 3388 (1975). }

\bibitem{25}  {\normalsize M.C.Wanke, A.G.Markelz, K.Unterrainer, S.J.Allen,
and R.Bhatt, {\it Proceedings of 23th International Conference on the
Physics of Semiconductors}, edited by N.Scheffter and R.Zimmerman, World
Scientific, Singapore (1996), p. 1791. }

\bibitem{26}  {\normalsize M. W. Feise and D. S. Citrin, Appl. Phys. Lett. 
{\bf 75}, 3536 (1999). }

\bibitem{???}  A.W.Ghosh, A.V.Kuznetsov, and J.W.Wilkins, Phys. Rev. Lett. 
{\bf 79}, 3494 (1997); E.P. Dodin, A.A. Zharov, and A.A. Ignatov, JETP {\bf %
87}, 1226 (1998).

\bibitem{27}  {\normalsize M. Holthaus, Z. Phys. B {\bf 89},251 (1992);
Phys. Rev. Lett. {\bf 69}, 351 (1992); M. Holthaus and D. Hone, Phys. Rev. B 
{\bf 47}, 6499 (1993). }

\bibitem{31}  Yu.A. Romanov and E.V. Demidov, Sov. Phys. Semicond. {\bf 31},
252 (1997).

\bibitem{32}  {\normalsize Yu.Yu. Romanova, E.V. Demidov, and Yu.A. Romanov,
Materials Science Forum {\bf 297-298}, 257 (1999); Yu.A. Romanov and E.V.
Demidov, Sov. Phys. Solid State {\bf 41}, 1555 (1999). }

\bibitem{41}  E.M. Conwell, {\it High Field Transport in Semiconductors},
Academic Press, New York and London, 1967.

\bibitem{42}  V.F. Gantmakher and Y.B. Levinson, {\it Carrier Scattering in
Metals and Semiconductors}, North Holland, Amsterdam, 1987.

\bibitem{28}  V.N. Bogomolov and T.M. Pavlova, Sov. Phys. Semicond. {\bf 29}%
, 428 (1995); B.N. Bogomolov, N.F. Kartenko, D.A. Kurdyukov, et al., Sov.
Phys. Solid State {\bf 41}, 313 (1999).

\bibitem{spr1}  G. Springholz, V. Holy, M. Pinczolits, and G. Bauer, Science 
{\bf 282}, 734 (1998).

\bibitem{spr2}  G. Springholz, M. Pinczolits, P. Mayer, V. Holy, G. Bauer,
H. H. Kang, and L. Salamanca-Riba, Phys. Rev. Lett. {\bf 84}, 4669 (2000).

\bibitem{29}  H. Lee, J.A. Johnson, M.Y. He, J.S.Speck, and P.M.Petroff,
Appl. Phys. Lett. {\bf 78}, 105 (2001).

\bibitem{Laz}  O.L. Lazarenkova and A.A. Balandin, J. Appl. Phys. {\bf 89},
5509 (2001).

\bibitem{34}  R.F.Kazarinov and R.A.Suris, Sov. Phys. Semicond. {\bf 6}, 148
(1972); {\it Ibid.} {\bf 7}, 488 (1973).

\bibitem{35}  {\normalsize J. Rotvig, A.-P. Jauho, and H. Smith, Phys. Rev.
Lett. {\bf 74}, 1831 (1995); Phys.Rev B {\bf 54}, 17691 (1996); W.-X. Yan,
X.-G. Zhao, and S.-Q. Bao, Physica B {\bf 252}, 63 (1998); M. Holthaus and
D.W. Hone, Phys. Rev. B {\bf 49}, 16605 (1994); E.Diez, R. Gomez-Alcala, F.
Dominguez-Adame, A. Sanchez, and G.P. Berman. Phys. Lett. A {\bf 240}, 109
(1998). }

\bibitem{39}  {\normalsize S.A.Ktitorov, G.S.Simin, and V.Ya.Sindalovsky.
Sov. Phys. Solid State {\bf 13}, 2230 (1971). }

\bibitem{40}  {\normalsize A.A. Ignatov, K.F. Renk, and E.P. Dodin, Phys.
Rev. Lett. {\bf 70}, 1996 (1993); K.N.Alekseev, G.P. Berman, D.K. Campbell,
E.H. Cannon, and M.C. Cargo. Phys. Rev. B, {\bf 54}, 10625 (1996);
K.N.Alekseev, E.H. Cannon, J.C. McKinney, F.V. Kusmartsev, and D.K.
Campbell. Phys. Rev. Lett. {\bf 80}, 2669 (1998); E.H. Cannon, F.V.
Kusmartsev, K.N. Alekseev, and D.K. Campbell. Phys. Rev. Lett. {\bf 85} 1302
(2000). }

\bibitem{33}  {\normalsize X.L. Lei, N.J.M. Horing, and H.L. Cui. Phys Rev.
Lett. {\bf 66}, 3277 (1991); J. Phys. Condens. Matter. {\bf 4}, 9375 (1992);
X.L. Lei and X.F. Wang. J.Appl. Phys. {\bf 73}, 3867 (1973); X.L. Lei,
N.J.M. Horing, H.L. Cui, and K.K. Thornber. Phys. Rev. B {\bf 48}, 5366
(1993); Solid State Commun. {\bf 86}, 231 (1993); Appl. Phys.Lett. {\bf 65},
2964 (1994). }

\bibitem{MS}  {\normalsize \ L.G.Mourokh and A.Yu.Smirnov,
J.Phys.:Condens.Matter {\bf 10}, 3213 (1998). }

\bibitem{Lei}  J.C. Cao, H.C. Liu, and X.L.Lei. Phys. Rev. B {\bf 61}, 5546
(2000).

\bibitem{44}  {\normalsize G. M. Shmelev and E.M. Epstein. Soviet Phys.
Solid St. {\bf 35}, 494 (1993). }

\bibitem{21}  Yu.A. Romanov and J.Yu. Romanova, JETP {\bf 91}, 1033 (2000).

\bibitem{45}  {\normalsize Yu.A. Romanov and L.K. Orlov, Sov. Phys. Solid
State {\bf 19}, 726 (1977). }
\end{references}
\end{document}